# Integrated time-synthetic photonic processor with programmable network topology

Bei Wu[1,2,3,4,#], Juzheng Li[1,2,3,4,#], Yucong Gao[1,2,3,4], Sen Wang[1,2,3,4], Yuhao Fang[5], Yaqi Li[1,2,3,4], Haiyao Luo[1,2,3,4], Li Zhang[1,2,3,4], William Shieh[5,6], Tao Chu[1,2,3,4], Hongsheng Chen[1,2,3,4,*], Bo Xiong[1,2,3,4,*] and Yihao Yang[1,2,3,4,*]

*[1] State Key Laboratory of Extreme Photonics and Instrumentation, ZJU-Hangzhou Global Scientific and Technological Innovation Center, Zhejiang University, Hangzhou 310027, China.*
*[2] International Joint Innovation Center, The Electromagnetics Academy at Zhejiang University, Zhejiang University, Haining 314400, China.*
*[3] Key Lab. of Advanced Micro/Nano Electronic Devices & Smart Systems of Zhejiang, Jinhua Institute of Zhejiang University, Zhejiang University, Jinhua 321099, China.*
*[4] Shaoxing Institute of Zhejiang University, Zhejiang University, Shaoxing 312000, China.*
*[5] School of Engineering, Westlake University, Hangzhou 310030, China.*
*[6] Zhejiang Key Laboratory of 3D Micro/Nano Fabrication and Characterization, Hangzhou 310030, China*

*[#]These authors contribute equally to this work.*
*[*]Corresponding authors: hansomchen@zju.edu.cn (H. Chen); bo.xiong@zju.edu.cn (B. Xiong); yangyihao@zju.edu.cn (Y. Yang)*

**Abstract**: Integrated photonic computing promises high-throughput, energy-efficient hardware to meet the surging computational demands of artificial intelligence (AI). However, most existing photonic processors encode information in spatial channels, where scaling requires adding physical components—escalating fabrication and calibration complexity—while fixed optical paths lock the network topology at fabrication. Here, we demonstrate an integrated photonic processor whose network topology is dynamically programmable in a time-synthetic dimension. Implemented on a thin-film lithium niobate (TFLN) chip, two coupled waveguide loops with a 97 ps round-trip time difference enable pulses in adjacent time slots to meet and interfere. These time slots serve as distinct lattice sites sharing the same physical coupler and phase modulators, reducing the hardware overhead of spatial scaling. By programming input pulse timing and loop phase shifts, we control both network connectivity and effective weights across successive round trips. On this photonic processor, we implement four convolution kernels for image edge detection, a recurrent neural network (RNN) for stock-price prediction with a 13.3 ns optical inference latency, and a double deep Q-network (DDQN) for blackjack decision-making. Our results establish temporal-dimension programming as a compact, post-fabrication reconfigurable paradigm for general-purpose photonic AI acceleration.

## Introduction

The rapidly expanding computational demands of AI have intensified the search for hardware that combines high throughput with low energy consumption. Integrated photonic circuits offer a compelling route to both, leveraging optical propagation and interference to execute massively parallel operations on a compact chip[1]. Architectures such as Mach-Zehnder interferometer (MZI) meshes[2,3], microring resonator arrays[4,5] and on-chip diffractive processors[6,7] have already demonstrated capabilities ranging from matrix multiplication to neural network inference. Yet most of these platforms encode information in spatial channels, distributing computation across cascaded physical components. As networks scale, device-to-device variations become increasingly difficult to calibrate, and interchannel crosstalk degrades fidelity. More fundamentally, spatial encoding binds the network architecture to the chip layout: although programmable weights enable diverse linear transformations, the available optical connectivity is fixed at fabrication, restricting the range of topologies a single chip can support.

Here we demonstrate an on-chip photonic processor whose network topology is dynamically programmable in a time-synthetic dimension. In time-synthetic photonics, the arrival time of optical pulses serves as a synthetic lattice coordinate—a principle previously explored for novel physics[8-17] and optical neural networks (ONNs)[18-20] in tabletop coupled-fiber loops. We translate this coupled-loop paradigm onto an integrated TFLN chip[21,22] comprising two unequal-length waveguide loops and two independently driven phase modulators. Because pulses occupying different time slots traverse the same physical coupler and modulators, the corresponding computational nodes share identical hardware. Scaling the network in time therefore mitigates the fabrication variations, interchannel crosstalk and per-element calibration overhead inherent to large spatial arrays. Within this fixed hardware, we program the network topology by controlling pulse injection and readout timing, and encode network weights by tailoring pulse interference across successive round trips. The on-chip TFLN modulators achieve an electro-optic bandwidth of 37.7 GHz and a half-wave voltage-length product of $V_\pi \cdot L$ ≈ 1.28 V·cm, enabling precise phase control of pulses spaced by 97 ps, while a low optical propagation loss of 0.32 dB/cm sustains multiple round trips.

Using this photonic processor, we implement three distinct computational paradigms: optical convolution, recurrent inference and deep reinforcement learning. For convolution, four optical pulses encode a 2 × 2 image patch and interfere over successive round trips; reading out a selected time slot yields the convolution output, with four programmed kernels extracting image edges of different orientations. To construct an RNN, we synchronize newly injected pulses with the returning pulses, predicting the next value in Tesla and Apple stock-price sequences from the preceding 16 samples with an optical inference latency of 13.3 ns. For reinforcement learning, we combine optical processing units (OPUs) in series and in parallel with residual connections to form the Q-value networks of a DDQN, which learns when to draw another card or stop in blackjack. These demonstrations establish integrated time-synthetic photonics as a compact, scalable and post-fabrication reconfigurable platform for general-purpose photonic computing.

## Results

### Photonic processor architecture

Artificial neurons distill biological neurons into static weighted interconnects followed by nonlinear activation, a simplification that discards the intrinsic temporal dynamics of biological information processing. As shown in Fig. 1a, a biological neuron's membrane potential evolves over time. Inspired by this dynamic behaviour, we develop a time-synthetic photonic processor on the TFLN platform.

The photonic processor comprises two unequal-length spiral waveguide loops coupled by a beamsplitter (Fig. 1b). During each round trip, optical pulses in the shorter loop arrive $\Delta t$ = 97 ps earlier than those in the longer loop. Pulses arriving simultaneously from the two loops interfere at the beamsplitter, redistributing their amplitudes during successive round trips. The relative temporal offset defines an effective spatial coordinate, while the round-trip number provides a temporal coordinate, mapping pulse propagation onto a 2D time-synthetic lattice.

The network topology is programmed by controlling pulse injection and readout timing, while the network weights are programmed through phase modulation. An electro-optic modulator (EOM) encodes input data in optical pulse amplitudes, which are then injected into the longer loop, the shorter loop or both, depending on the task. Independently driven TFLN phase modulators control the phase shifts in each loop, thereby governing their subsequent interference. Under ideal lossless conditions, pulse dynamics obey the modified discrete quantum walk equations[23],

$$u_{m-1}^{n} = [u_{m}^{n-1} + i v_{m}^{n-1}] e^{i\varphi_u(m,\, n)}/\sqrt{2} \tag{1}$$

$$v_{m+1}^{n} = [i u_{m}^{n-1} + v_{m}^{n-1}] e^{i\varphi_v(m,\, n)}/\sqrt{2} \tag{2}$$

where $u_m^n/(v_m^n)$ denote the complex pulse amplitudes at lattice position $m$ and round trip $n$ in the shorter/(longer) loop. The corresponding phase shifts, $\varphi_u(m, n)$ and $\varphi_v(m, n)$, are programmable over $[0,\, 2\pi)$. Together, the selected input time slots and programmed phase sequences determine the effective complex-valued weight matrix, allowing different neural networks to be implemented without modifying the physical architecture. Specifically, we configure the chip to perform optical convolution for image edge detection, RNN inference for stock price prediction and DDQN decision-making for blackjack gameplay (Fig. 1c-e).

To maintain signal quality, we monitor the optical signals in real time and, when the signal-to-noise ratio falls below a predefined threshold, re-encode the measured outputs onto fresh optical pulses and reinject them into the chip for subsequent processing. This approach is commonly used in integrated photonic computing to support large-scale ONNs[5,6,24,25]. To further extend the network depth, we propose an erbium-doped TFLN implementation in which optically pumped spiral waveguides provide on-chip gain to compensate for propagation loss[26]; see Supplementary Note 6 for details.

**Photonic chip characterization**

We fabricate the photonic chip on a commercial X-cut TFLN-on-insulator wafer (NANOLN), as shown in Fig. 2a. Each phase modulator uses a single optical waveguide passing through the signal-ground gap of a coplanar transmission line (Fig. 2b). The electrodes are separated from the optical waveguide by a $SiO_2$ cladding layer (Fig. 2c), minimizing the interaction between the optical field and the metal electrodes and thereby reducing optical propagation loss[27]. The microwave transmission lines are terminated with on-chip titanium/platinum (Ti/Pt) resistors for impedance matching, thereby suppressing microwave reflections (Fig. 2d, left). To characterize the phase modulators, an MZI consisting of two phase modulators fabricated on the same chip is initially tested[28], with its schematic shown in Supplementary Fig. 6. Specifically, a 4 mm-long modulator exhibits an electro-optic bandwidth of 37.7 GHz, an electrical reflection coefficient below -10 dB, and a half-wave voltage-length product ($V_\pi \cdot L$) of approximately 1.28 V·cm (Fig. 2e,f). The spiral waveguides have a minimum bend radius of 100 μm, with adjacent waveguides separated by a 10 μm gap to suppress coupling. Optimized reactive ion etching produces smooth waveguide sidewalls (Fig. 2d, right), yielding a low optical propagation loss of 0.32 dB/cm.

The modulator's high-speed electro-optic response enables rapid control of closely spaced optical pulses, supporting high computational throughput in the photonic chip. As described by Eqs. 1 and 2, each modulation interval involves two complex-valued multiplications, corresponding to eight real-valued multiplications and four real-valued additions[5]. Hence, a single computing core thus provides a theoretical computational throughput of $R$ = 12 operations × 37.7 GHz ≈ 452 GOPS. In addition, the photonic chip has an energy consumption of 450 fJ/OP, corresponding to an energy efficiency of 2.21 tera-operations per second per watt (TOPS/W); see Supplementary Note 4 for details. This computational throughput can be further improved by leveraging mature parallelization technologies from optical communications, including wavelength-division multiplexing and multimode waveguides[29]. Details of the proposed wavelength-division multiplexing scheme are provided in Supplementary Note 7.

**Image edge detection with optical convolution**

We demonstrate programmable optical convolution for image edge detection, as illustrated in Fig. 3a. The input image is flattened into a sequence of pixel values; each group of four consecutive values corresponds to a 2 × 2 pixel block. An EOM encodes these values in 60 ps-wide optical pulses spaced 97 ps apart. The pulses are then injected into the longer loop and occupy four adjacent sites in the time-synthetic lattice.

Given the input vector $\boldsymbol{X} = (x_1, x_2, x_3, x_4)^{\mathrm{T}}$, after six round trips we take the third pulse in the longer loop as the output. Its amplitude is $y = \boldsymbol{WX} = \boldsymbol{W_6W_5W_4W_3W_2W_1X} = \sum_{j=1}^{4} w_j x_j$, and the corresponding optical intensity is $|y|^2$. Here, $\boldsymbol{W_n}$ is the transfer matrix for the $n$-th round trip, and $\boldsymbol{W} = (w_1, w_2, w_3, w_4) \in \mathbb{C}^{1\times 4}$ is the overall effective transfer matrix from the input to the selected output; see Supplementary Note 1 for details. The four entries of $\boldsymbol{W}$ are the weights of a convolution kernel and can be tuned by programming the phase sequences.

For each target kernel, the phase sequences are optimized by backpropagation, and Fig. 3b shows a representative numerical optimization, where the mean squared error (MSE) between the optimized and target convolution kernel decreases to $7.6 \times 10^{-13}$. As shown in Fig. 3c, the measured results (blue lines) are in good agreement with expectations (black lines). The optical convolution runtime is 4.211 ns per 2 × 2 image patch, over four orders of magnitude shorter than the 97 μs required on a conventional central processing unit (CPU). Figure 3d shows that different kernels extract distinct image edges, demonstrating the programmability of our photonic chip.

**Stock price prediction with optical RNN**

We next configure the time-synthetic photonic processor as an optical RNN for sequential prediction. As in the multilayer electronic RNN shown in Fig. 4a, which combines feedforward connections between hidden layers with recurrent connections that carry information from previous time steps, the optical RNN feeds the output pulses back to the input port after each round trip; they meet and interfere with the newly injected input pulses. At time step $t$, the hidden state of layer $l$ depends on the hidden state of layer $l-1$ at time step $t$ and that of layer $l$ at time $t-1$: $h_t^l = w_{hx}^l h_t^{l-1} + w_{hh}^l h_{t-1}^l$, where $h_t^0 = x_t$. For a network with $L$ hidden layers, the output is $o_t = w_{oh}^L h_t^L$. The input, recurrent and output weights $w_{hx}^l$, $w_{hh}^l$ and $w_{oh}^L$, respectively, are determined by the phase sequences in the longer and shorter loops; see Supplementary Note 2 for details. Two phase modulators are reused throughout the computation; therefore, the photonic processor is less susceptible to fabrication and calibration errors, making it well suited for high-precision analogue computation.

We use the optical RNN to predict the next stock price from the preceding 16 samples. The normalized sequence is divided into four groups of samples, which are processed by optical processing units (OPU1 - OPU4). Each OPU processes four input pulses over seven round trips. The input pulses have a width of 60 ps and a spacing of 621 ps, which equals the shorter loop period to ensure newly injected pulses meet and interfere with the returning pulses. For each subsequent OPU, the input pulse amplitudes are given by the sum of the new input signals and complex output amplitudes of the preceding OPU. After OPU4, the magnitude of the final pulse amplitude, $|y_4|$, gives the normalized prediction, which is then converted back to the original price scale.

Stock prices are sampled at one-minute intervals, with prices from 08:00 to 20:00 UTC used for training and prices from 20:00 to 23:59 UTC used for testing. Figures 4b and 4c show the predicted and actual stock prices for Tesla and Apple on 1 October 2026. The predictions closely follow the overall price trajectories in both the training and test sets, with modest local deviations and a slight time delay due to model approximation and experimental imperfections, including propagation loss and detection noise. Future improvements in network architecture and training strategies could enhance prediction accuracy, while low-loss waveguides, on-chip optical amplification and low-noise photodetectors could further improve signal quality. The optical prediction runtime is 13.3 ns, approximately five orders of magnitude shorter than the 1.3 ms required on a conventional CPU. Such low latency could reduce computational delays in high-frequency trading, enabling more timely trading decisions.

**Decision-making with optical DDQN**

We further configure the time-synthetic photonic processor to implement a DDQN, with online and target Q-networks sharing the same architecture in Fig. 5a. Each Q-network combines three sequential OPUs with residual connections and three parallel OPUs, with optical pulses injected into the longer loop and each OPU processing three input pulses over four round trips (Fig. 5b). Details of the Q-network architecture are provided in Supplementary Note 3.

We use this architecture to learn a policy for playing blackjack (Fig. 5c), with the game rules described in Methods. The game state is described as $s = (p, c, u)$, where $p \in \{4, 5, \ldots, 21\}$ is the player's total card, $c \in \{1, 2, \ldots, 10\}$ is the dealer's visible card and $u \in \{0, 1\}$ indicates whether the player holds a usable ace. The game state is input into the Q-network, and the outputs $y_2$ and $y_3$ estimate the Q-values for stick and hit. During training, the agent interacts with the blackjack environment and stores each transition $(s_i, a_i, s_i', r_i, d_i)$ in a replay memory, with each element denoting the current state, action, next state, reward and game termination flag, respectively. Minibatches sampled from this memory are used to train the online Q-network.

DDQN reduces Q-value overestimation by using one network to select the action and another to evaluate it. Specifically, the online Q-network selects an action according to $a_i^* = \mathrm{argmax}_{a\in\mathcal{A}} Q(s_i', a; \theta)$, and the target Q-network evaluates this action; it computes the temporal-difference (TD) target $y_i = r_i + \gamma(1 - d_i)Q(s_i', a_i^*; \theta^-)$. Here, $\mathcal{A}$ = {stick, hit} is the action set, $\theta$ and $\theta^-$ are the parameters of the online and target Q-networks, respectively, and $\gamma$ is the discount factor. The online Q-network is optimized by minimizing the smooth L1 loss between its predicted Q-values and the TD targets, and the target Q-network parameters are periodically copied from the online Q-network, $\theta^- \leftarrow \theta$.

Figure 5d shows that the mean reward increases from approximately -0.424 to -0.05 over $10^5$ training episodes. Even the optimal policy has a negative expected reward because the game rules favor the dealer: if the player busts, the dealer wins immediately without taking any action. Figures 5e and 5f show the learned policies and predicted Q-values across the player's total card and the dealer's visible card. When the player has a usable ace, the learned policy is more inclined to hit, since an ace can count as either 1 or 11 and thus reduces the risk of busting. For random game states in Fig. 5g, the online Q-network selects the action with the higher Q-value; the lower panels show the optical intensity distributions obtained by coherently summing the complex amplitudes from the final three parallel OPUs, with the intensities of output pulses $y_2$ and $y_3$ representing stick and hit, respectively. These results show that optical Q-networks can learn to make decisions through interaction with the environment.

## Discussion

In this work, we design and fabricate an integrated time-synthetic photonic processor on the TFLN platform, which processes sequential optical pulses with a symbol period of 97 ps and a symbol rate of 10.3 Gbaud. By mapping computational nodes onto distinct time slots that share the same physical coupler and phase modulators, our photonic processor mitigates the device-to-device variations, interchannel crosstalk and calibration overhead associated with scaling spatial arrays[30,31]. Within a fixed hardware, the network topology and weights can be programmed by controlling pulse injection and readout timing, and tailoring pulse interference across successive round trips. The implementation of optical convolution, recurrent inference and deep reinforcement learning demonstrates the ability of the photonic processor to support distinct computational architectures after fabrication. These results establish integrated time-synthetic photonics as a compact, reconfigurable platform for general-purpose photonic AI acceleration with subnanosecond modulation intervals.

Substantial improvements are feasible in the near future. For example, higher electro-optic bandwidth could enable faster computation; wavelength-division multiplexing could increase throughput by processing signals in parallel; and improving the electro-optic modulation efficiency would reduce the electrical energy required for high-speed electro-optic control. To enable longer input sequences and deeper networks within a single OPU, low-loss waveguides and low-noise on-chip amplification[26,32-33] could support pulse propagation over more round trips. At the system level, integrating laser sources and photodetectors with the photonic processor could further combine optical generation, processing and readout in a single compact module[21]. These directions highlight the compatibility of our integrated time-synthetic architecture with established photonic technologies and its potential as a general-purpose, high-performance module for AI acceleration.

## Methods

### Device fabrication

The devices in our study are fabricated using a commercial X-cut TFLN-on-insulator wafer (NANOLN). The wafer comprises a 400 nm-thick LN layer and a 4.7 µm-thick buried oxide layer (thermally grown) on a 500 µm-thick silicon (Si) handle. Figure 2c shows the cross-section of the TFLN phase modulator, and Table 1 lists the corresponding key design parameters. The fabrication process involves the following steps. (1) Patterning of the optical layer: Stepper lithography is used to define the waveguide patterns, which are then transferred into the TFLN layer by $Ar^+$-based inductively coupled plasma (ICP) dry etching to a depth of 200 nm, forming rib waveguides. (2) Cladding with silicon dioxide ($SiO_2$): An 800 nm-thick $SiO_2$ layer is deposited over the devices using plasma-enhanced chemical vapour deposition (PECVD). The cladding is then selectively etched to a depth of 600 nm using ICP dry etching to form recesses, leaving 200 nm of $SiO_2$ beneath the electrodes. (3) Depositing the Ti/Pt resistor: The Ti/Pt layer is defined using photolithography with deposition through electron-beam evaporation, followed by a bilayer lift-off process. (4) Defining the double-layer electrode: Microwave electrodes are defined using a combination of photolithography, electron-beam evaporation, and a bilayer lift-off process. Using this approach, a 600 nm-thick layer of gold (Au) and a 20 nm-thick layer of Ti are deposited to form the microwave electrodes. The fabrication process results in a photonic chip with high-quality optical components and a well-defined microwave circuit, enabling high-performance computing operations on TFLN.

**Experimental architecture and setup**

Continuous-wave light at 1550 nm passes through a polarization controller (PC) and an EOM (AFR AM40), which generates optical pulses and encodes the input vector in their amplitudes. The pulses are amplified by an erbium-doped fiber amplifier (EDFA), pass through a second PC and are injected into the longer loop of the photonic chip. Each loop contains a TFLN phase modulator driven in synchrony with the optical pulses. The electrical drive waveforms for the EOM and the two on-chip phase modulators are generated by an arbitrary waveform generator (AWG, Keysight M8194A) operating at a sampling rate of 120 GSa/s with four independent channels. The phase modulators apply the programmable phase sequences $e^{i\varphi_u}$ and $e^{i\varphi_v}$ in the shorter and longer loops, respectively, with the ideal lossless pulse evolution described by Eqs. 1 and 2. At each round trip, 10% of the optical power is tapped from the photonic chip for readout, while the remaining light continues to propagate in the loops. The tapped pulses are amplified by another EDFA, filtered by a bandpass filter (BPF) and coherently detected to recover their complex amplitudes; see Supplementary Note 5 for details. The resulting electrical waveforms are recorded using a real-time oscilloscope (Keysight UXR0594A) at a sampling rate of 256 GSa/s.

**Image sources**

The four images used for optical convolution (Fig. 3d) are, from left to right, *Girl with a Pearl Earring* by Johannes Vermeer, *The Goldfinch* by Carel Fabritius, *Man with a Glove* by Titian, and *Portrait of Madame X* by John Singer Sargent.

**Blackjack's game rules**

In blackjack, the player aims to beat the dealer without exceeding a total card of 21. At the start of each game, both receive two cards, with one of the dealer's cards hidden from the player. Number cards (2 - 10) take their face values, whereas jacks (J), queens (Q) and kings (K) each count as 10. An ace (A) can count as either 1 or 11. It is counted as 11 whenever this keeps the hand total at or below 21, and as 1 otherwise. The player repeatedly chooses either to draw another card (hit) or stop drawing (stick). If the player's total exceeds 21, the player busts and loses immediately; otherwise, the player can make another decision. Once the player sticks, the dealer draws additional cards until the total is 17 or higher. If the dealer exceeds 21, the player wins; otherwise, the higher total wins, and equal totals result in a draw. In the optical DDQN, the player acts as the agent. The rewards are encoded from the player's perspective: +1 for a win, -1 for a loss and 0 for a draw. A hit that does not terminate the game also receives a reward of 0. These rewards provide the feedback to train the online Q-network.

**Table 1|Design parameters of the TFLN phase modulator.**

| Parameter | Value |
|---|---|
| w | 900 nm |
| t | 200 nm |
| $t_0$ | 200 nm |
| $t_1$ | 4.7 µm |
| $t_2$ | 500 µm |
| $h_1$ | 800 nm |
| $h_2$ | 600 nm |

**Data availability**

All data needed to evaluate the conclusions in the paper are present in the main text or the supplementary materials.

**Code availability**

The code used for simulation and data plotting is available from the corresponding authors upon request.

**References**

1. Bogaerts, W., et al. Programmable photonic circuits. *Nature* **586**, 207-216 (2020).
2. Shen, Y., et al. Deep learning with coherent nanophotonic circuits. *Nat. Photon.* **11**, 441-446 (2017).
3. Bandyopadhyay, S., et al. Single-chip photonic deep neural network with forward-only training. *Nat. Photon.* **18**, 1335-1343 (2024).
4. Xiao, Z., et al. End-to-end all-optical in-sensor computing system using photonic integrated circuits. *Sci. Adv.* **12**, eaef8657 (2026).

5. Yan, T., et al. A complete photonic integrated neuron for nonlinear all-optical computing. *Nat. Comput. Sci.* **5**, 1202-1213 (2025).
6. Wang, C., et al. Diffractive tensorized unit for million-TOPS general-purpose computing. *Nat. Photon.* **19**, 1078-1087 (2025).
7. Chen, C., et al. High-precision integrated diffractive optical networks enabling regression analysis. *Nat. Commun.* **17**, 9870 (2026).
8. Ozawa, T. & Price, H. M. Topological quantum matter in synthetic dimensions. *Nat. Rev. Phys.* **1**, 349-357 (2019).
9. Weidemann, S., et al. Topological funneling of light. *Science* **368**, 311-314 (2020).
10. Weidemann, S., et al. Coexistence of dynamical delocalization and spectral localization through stochastic dissipation. *Nat. Photon.* **15**, 576-581 (2021).
11. Weidemann, S., Kremer, M., Longhi, S. & Szameit, A. Topological triple phase transition in non-Hermitian Floquet quasicrystals. *Nature* **601**, 354-359 (2022).
12. Wang, S., et al. High-order dynamic localization and tunable temporal cloaking in ac-electric-field driven synthetic lattices. *Nat. Commun.* **13**, 7653 (2022).
13. Marques Muniz, A. L., et al. Observation of photon-photon thermodynamic processes under negative optical temperature conditions. *Science* **379**, 1019-1023 (2023).
14. Qin, C., et al. Observation of discrete-light temporal refraction by moving potentials with broken Galilean invariance. *Nat. Commun.* **15**, 5444 (2024).
15. Feis, J., et al. Space-time-topological events in photonic quantum walks. *Nat. Photon*. **19**, 518-525 (2025).
16. Huang, H., Du, Z., Liao, K., Hu, X. & Gong, Q. Reconfigurable high-dimensional synthetic photonic lattices. *Sci. Adv.* **11**, eadw7198 (2025).
17. Ren, Y., et al. Observation of momentum-gap topology of light at temporal interfaces in a time-synthetic lattice. *Nat. Commun.* **16**, 707 (2025).
18. Pankov, A. V., et al. Optical neural network based on synthetic nonlinear photonic lattices. *Phys. Rev. Appl.* **17**, 024011 (2022).
19. Peng, B., et al. Optical neural network architecture for deep learning with temporal synthetic dimension. *Chin. Phys. Lett*. **40**, 034201 (2023).
20. Wu, B., et al. Time-synthetic optical neural networks with stable programmable gain. *Nat. Commun.* **17**, 6068 (2026).
21. Hu, Y., et al. Integrated lithium niobate photonic computing circuit based on efficient and high-speed electro-optic conversion. *Nat. Commun.* **16**, 8178 (2025).
22. Hu, Y., et al. Integrated electro-optics on thin-film lithium niobate. *Nat. Rev. Phys*. **7**, 237-254 (2025).
23. Schreiber, A., et al. Photons walking the line: a quantum walk with adjustable coin operations. *Phys. Rev. Lett.* **104**, 050502 (2010).
24. Xu, Z., et al. Large-scale photonic chiplet Taichi empowers 160-TOPS/W artificial general intelligence. *Science* **384**, 202-209 (2024).
25. Wang, D., et al. Ultrafast silicon photonic reservoir computing engine delivering over 200 TOPS. *Nat. Commun.* **15**, 10841 (2024).
26. Wang, Y., et al. Unifying optical gain and electro-optical dynamics in Er-doped thin-film lithium niobate platform. *Nat. Commun.* **16**, 10462 (2025).
27. Xu, M., et al. Dual-polarization thin-film lithium niobate in-phase quadrature modulators for terabit-per-second transmission. *Optica* **9**, 61-62 (2022).
28. Xie, Y., et al. Ultra-wideband tunable thin-film lithium-niobate-on-insulator microwave photonic filter. *ACS Photon.* **12**, 1689-1697 (2025).

29. Zou, K., et al. High-capacity free-space optical communications using wavelength- and mode-division-multiplexing in the mid-infrared region. *Nat. Commun.* **13**, 7662 (2022).
30. Reck, M., Zeilinger, A., Bernstein, H. J. & Bertani, P. Experimental realization of any discrete unitary operator. *Phys. Rev. Lett.* **73**, 58 (1994).
31. Clements, W. R., Humphreys, P. C., Metcalf, B. J., Kolthammer, W. S. & Walmsley, I. A. Optimal design for universal multiport interferometers. *Optica* **3**, 1460-1465 (2016).
32. Liang, Y., et al. A high-gain cladded waveguide amplifier on erbium doped thin-film lithium niobate fabricated using photolithography assisted chemo-mechanical etching. *Nanophotonics* **11**, 1033-1040 (2022).
33. Yang, L., et al. A photonic integrated circuit-based erbium-doped amplifier. *Science* **376**, 1309-1313 (2022).

**Acknowledgments**

Y.Y. discloses support for the research of this work from the Key Research and Development Program of the Ministry of Science and Technology [grant numbers 2022YFA1405200 and 2022YFA1404900], the Fundamental Research Funds for the Zhejiang Provincial Universities [grant number 226-2025-00231], the Science Challenge Project [grant number TZ2025015] and the National Natural Science Foundation of China [grant number U25D8017]. H.C. discloses support for the research of this work from the Key Research and Development Program of the Ministry of Science and Technology [grant number 2022YFA1404704].

**Author Contributions Statement**

Y.Y. and B.W. conceived the idea of this research. B.W. proposed the algorithm structure. J.L., Y.G. and S.W. fabricated the photonic chip. Y.F., S.W., J.L., Y.G. and B.W. conducted the experiment. Y.Y. and B.W. wrote the paper. All authors shared their insights and contributed to discussions on the results. H.C., B.X. and Y.Y. supervised the project.

**Competing Interests Statement**

The authors declare no competing interests.

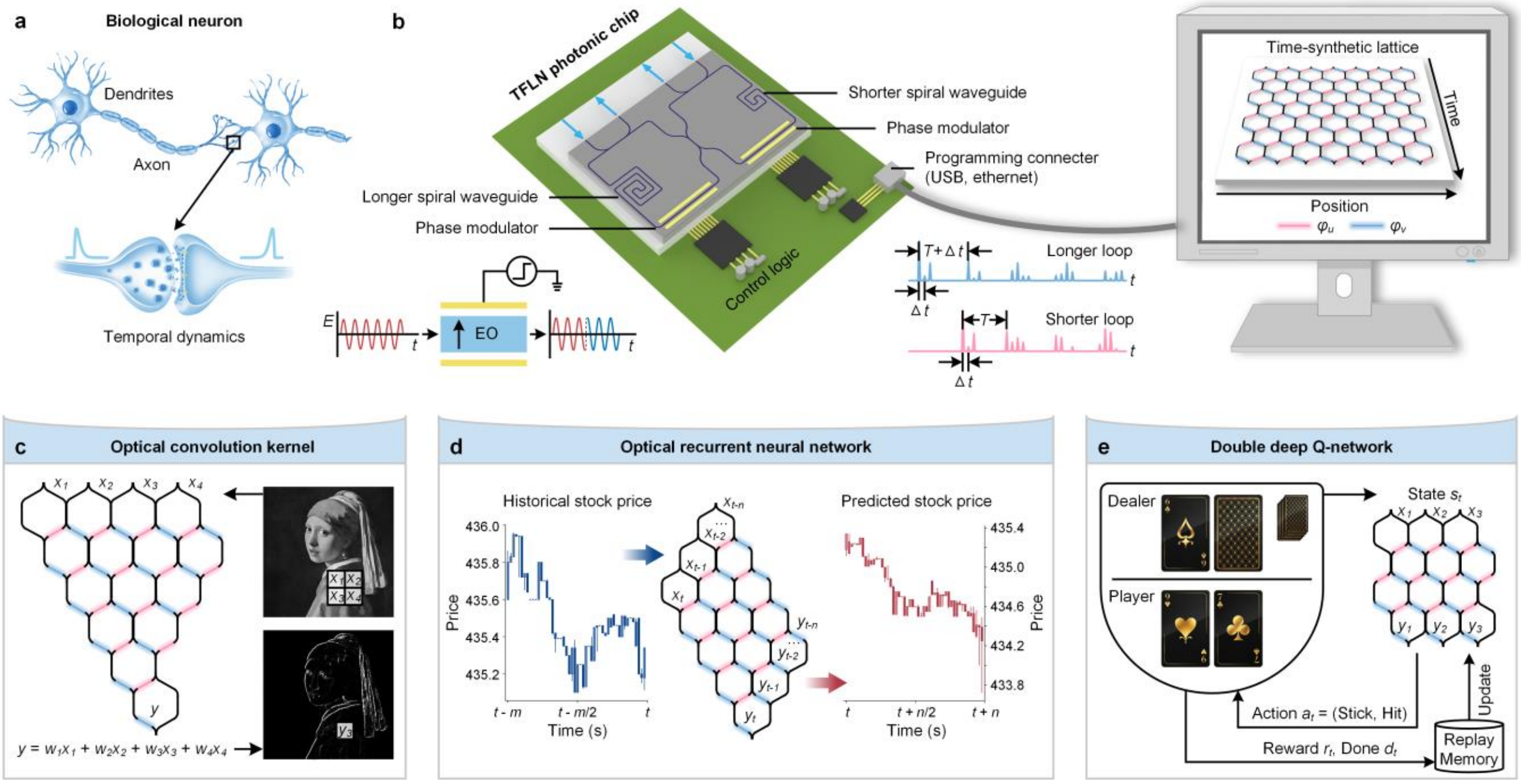

**Fig. 1 | A time-synthetic photonic processor on the TFLN platform. a,** Temporal dynamics of biological neurons. **b,** Schematic of the photonic computing chip and its mapping onto the time-synthetic lattice. **c-e,** Applications of the photonic processor to optical convolution for image edge detection (**c**), RNNs for stock price prediction (**d**) and DDQN for blackjack gameplay (**e**). TFLN thin-film lithium niobate, USB universal serial bus, EO electro-optic.

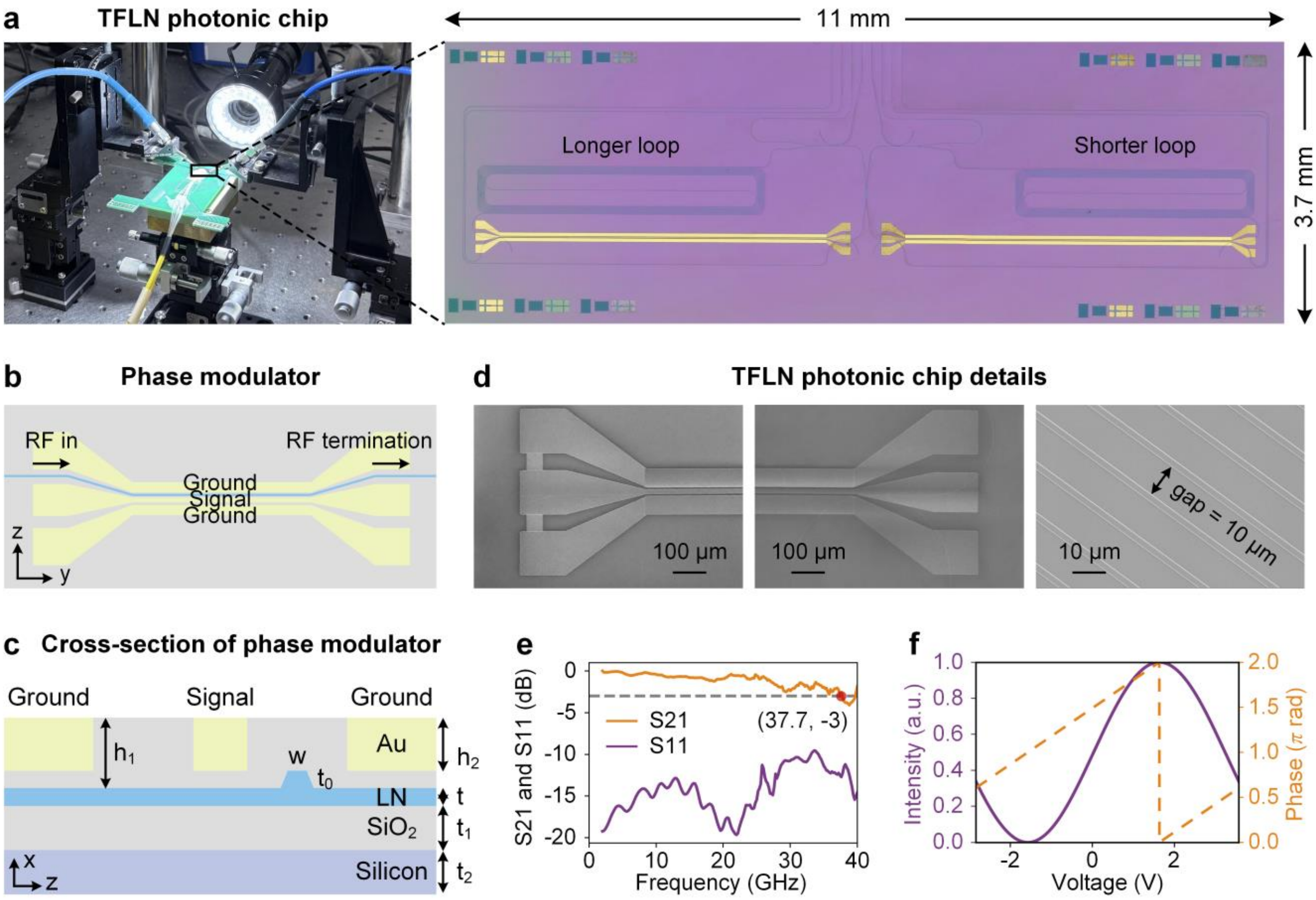

**Fig. 2 | Integrated TFLN photonic chip implementation. a,** Left panel, packaged photonic chip with an optical fiber array. Right panel, optical microscope image of the fabricated photonic chip. **b,c,** Schematic (**b**) and cross-section illustration (**c**) of the TFLN phase modulator. **d,** Scanning electron microscope images of phase modulator details and high-quality waveguides. **e,** Electro-optic forward transmission ($S_{21}$ EO) and electric-electric input reflection ($S_{11}$ EE) response of a 4 mm-long TFLN modulator. **f,** Electro-optic modulation efficiency of a 4 mm-long TFLN modulator, with a $V_{\pi}{\cdot}L$ of 1.28 V·cm.

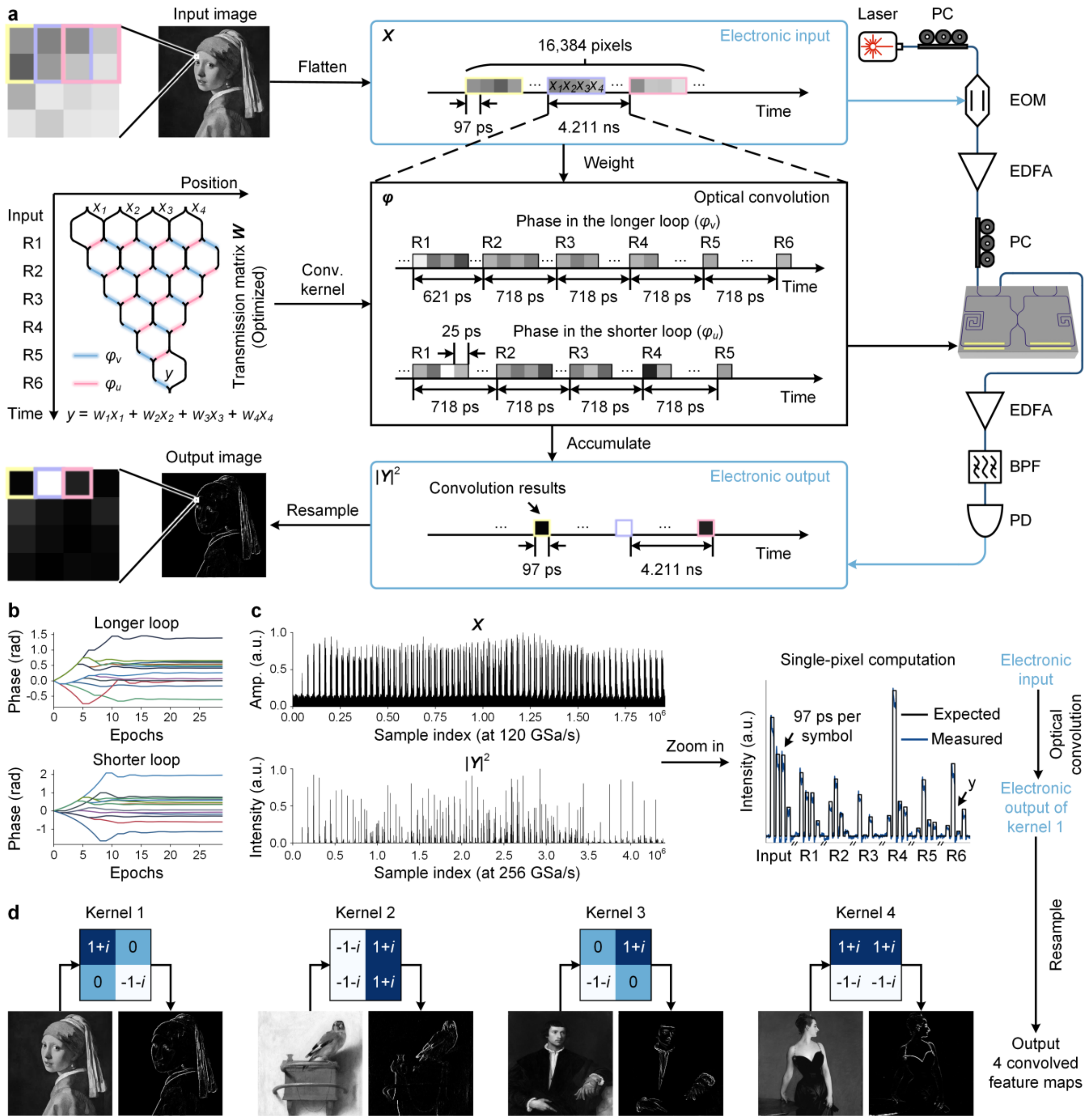

**Fig. 3 | Programmable optical convolution for image edge detection. a,** Left panel, data-processing pipeline. Right panel, simplified experimental setup. **b,** Evolution of phase sequences during optimization. **c,** Upper left, input pulse amplitudes that encode the flattened input image. Lower left, output pulse intensities that represent the convolution results. Right, comparison of measured and expected pulse intensities for a representative single-pixel computation. **d,** Input images and corresponding outputs obtained using four convolution kernels designed to detect -45°, right, 45° and upper edges. PC polarization controller, EOM electro-optic modulator, EDFA erbium-doped fiber amplifier, BPF bandpass filter, PD photodetector, Amp. amplitude.

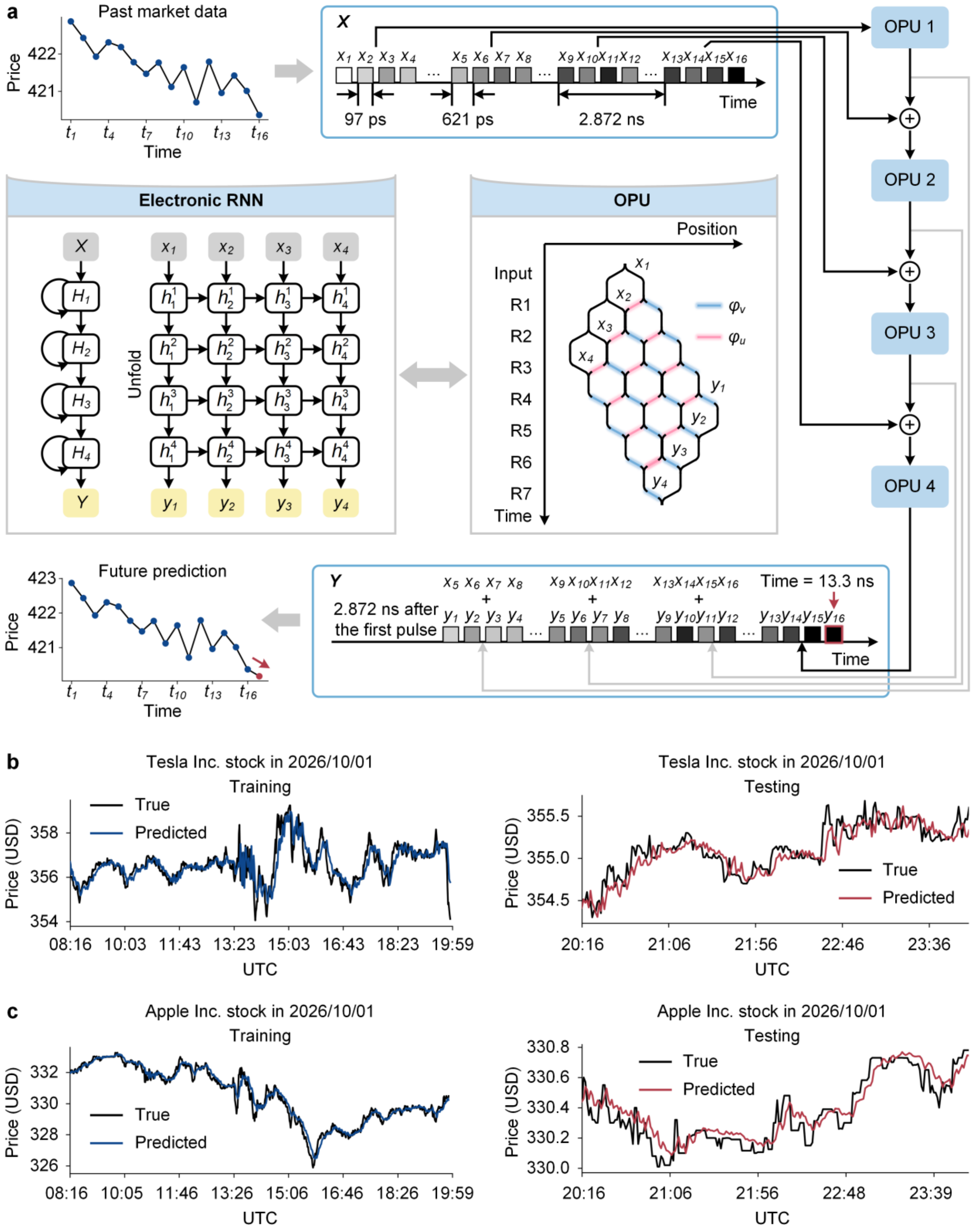

**Fig. 4 | Ultrafast optical RNN for stock price prediction. a,** Optical RNN architecture and data-processing pipeline, showing the structural similarity between an unfolded electronic RNN and an OPU. **b,c,** Actual and predicted stock prices for Tesla (**b**) and Apple (**c**) on 1 October 2026, with training results on the left and test results on the right. OPU optical processing unit, RNN recurrent neural network, Inc. incorporated, UTC coordinated universal time.

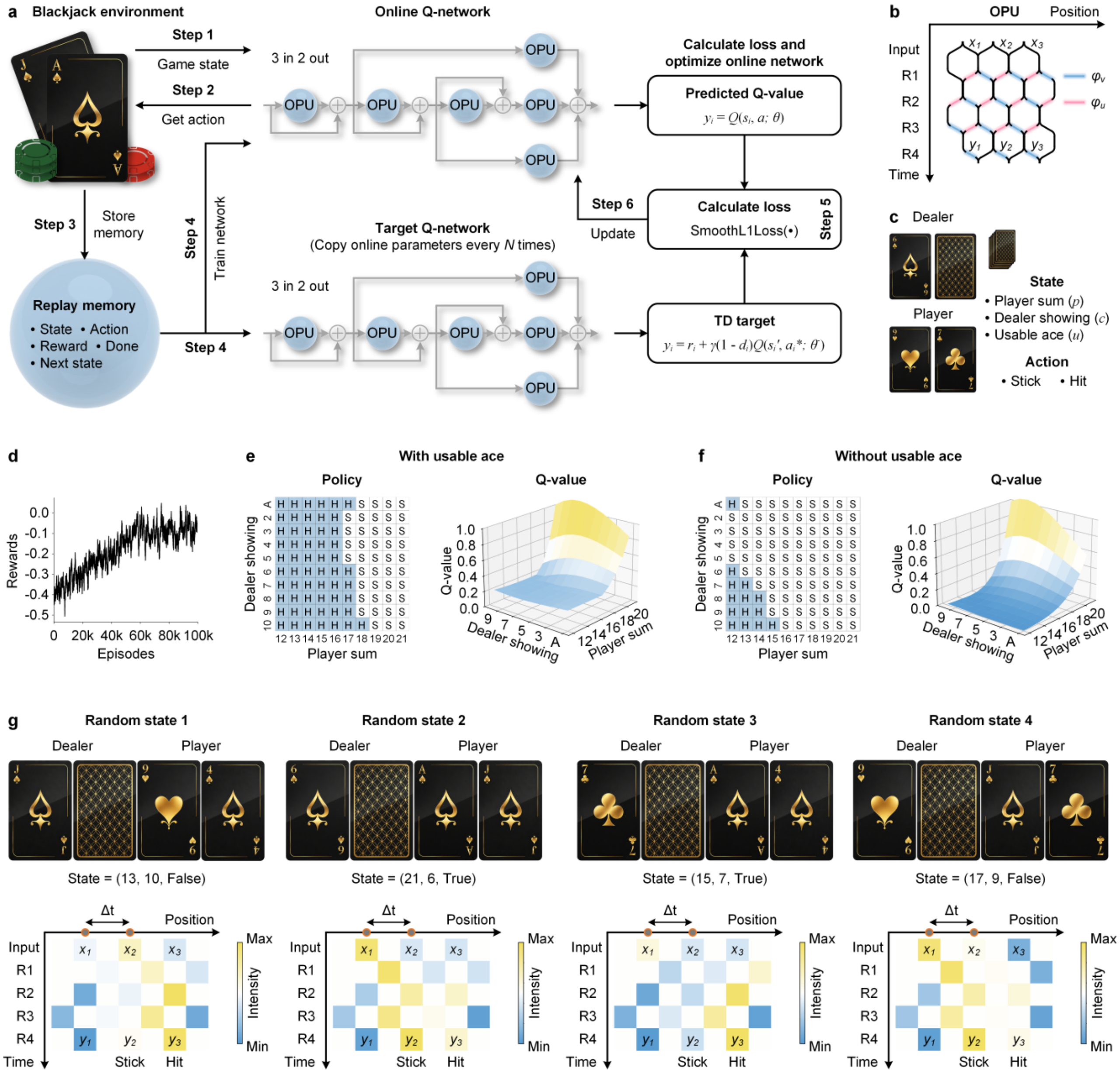

**Fig. 5 | Optical DDQN for decision-making. a,** Optical DDQN network architecture and training workflow. **b,** Evolution of three input pulses over four round trips in an OPU. **c,** Blackjack environment, showing the game states and available actions. **d,** Evolution of the mean reward during training. **e, f,** Learned policies (left) and predicted Q-values (right) for game states with (**e**) and without (**f**) a usable ace. H and S denote hit and stick, respectively. **g,** Top panel, four random game states. Bottom panel, corresponding optical intensity distributions during propagation. Intensities of output pulses $y_2$ and $y_3$ correspond to stick and hit, respectively.